\documentclass[8.5pt,twoside,twocolumn]{article}
\usepackage[super,sort&compress,comma]{natbib}
\usepackage{mhchem}
\usepackage{times,mathptmx}
\usepackage{sectsty,balance,url,graphicx,lastpage,amssymb}
\usepackage[format=plain,justification=raggedright,singlelinecheck=false,font=small,labelfont=bf,labelsep=space]{caption}
\usepackage{fancyhdr}
\begin{document}

\thispagestyle{plain}
\fancypagestyle{plain}{
\fancyhead[L]{\textbf{PAPER}}
\renewcommand{\headrulewidth}{1pt}}
\renewcommand{\thefootnote}{\fnsymbol{footnote}}
\renewcommand\footnoterule{\vspace*{1pt}%
\hrule width 3.4in height 0.4pt \vspace*{5pt}}
\setcounter{secnumdepth}{5}

\makeatletter
\def\subsubsection{\@startsection{subsubsection}{3}{10pt}{-1.25ex plus -1ex minus -.1ex}{0ex plus 0ex}{\normalsize\bf}}
\def\paragraph{\@startsection{paragraph}{4}{10pt}{-1.25ex plus -1ex minus -.1ex}{0ex plus 0ex}{\normalsize\textit}}
\renewcommand\@biblabel[1]{#1}
\renewcommand\@makefntext[1]%
{\noindent\makebox[0pt][r]{\@thefnmark\,}#1}
\makeatother
\renewcommand{\figurename}{\small{Fig.}~}
\sectionfont{\large}
\subsectionfont{\normalsize}

\fancyfoot{}
\fancyfoot[RO]{\footnotesize{\sffamily{1--\pageref{LastPage} ~\textbar  \hspace{2pt}\thepage}}}
\fancyfoot[LE]{\footnotesize{\sffamily{\thepage~\textbar\hspace{3.45cm} 1--\pageref{LastPage}}}}
\fancyhead{}
\renewcommand{\headrulewidth}{1pt}
\renewcommand{\footrulewidth}{1pt}
\setlength{\arrayrulewidth}{1pt}
\setlength{\columnsep}{6.5mm}
\setlength\bibsep{1pt}

\twocolumn[
  \begin{@twocolumnfalse}
\noindent\LARGE{\textbf{Exploring PtSO$_4$ and PdSO$_4$ phases: an evolutionary algorithm based investigation}}
\vspace{0.6cm}

\noindent\large{\textbf{Hom Sharma,\textit{$^{a}$} Vinit Sharma,\textit{$^{b}$} and
Tran Doan Huan$^{\ast,}$\textit{$^{b,c}$}}}\vspace{0.5cm}


\vspace{0.6cm}

\noindent \normalsize{Metal sulfate formation is one of the major challenges to the emissions aftertreatment catalysts. Unlike the incredibly sulfation prone nature of Pd to form PdSO$_4$, no experimental evidence exits for the PtSO$_4$ formation. Given the mystery of nonexistence of the PtSO$_4$, we explore the PtSO$_4$ using a combined approach of evolutionary algorithm based search technique and quantum mechanical computations. Experimentally known PdSO$_4$ is considered for the comparison and validation of our results. We predict many possible low-energy phases of the PtSO$_4$ and PdSO$_4$ at 0K, which are further investigated under wide range of temperature-pressure conditions. An entirely new low-energy (tetragonal $P4_2/m$) structure of the PtSO$_4$ and PdSO$_4$ is predicted, which appears to be the most stable phase of the PtSO$_4$ and a competing phase of the experimentally known monoclinic $C_2/c$ phase of PdSO$_4$. Phase stability at finite temperature is further examined and verified by Gibbs free energy calculations of sulfates towards their possible decomposition products. Finally, temperature-pressure phase diagrams are computationally established for both PtSO$_4$ and PdSO$_4$. }
\vspace{0.5cm}
 \end{@twocolumnfalse}
  ]



\footnotetext{\textit{$^{a}$~Department of Chemical and Biomolecular Engineering, University of Connecticut, Storrs, CT 06269 USA.}}
\footnotetext{\textit{$^{b}$~Materials Science and Engineering, University of Connecticut, Storrs, CT 06269 USA }}
\footnotetext{\textit{$^{c}$~Institute of Engineering Physics, Hanoi University of Science and Technology, 1 Dai Co Viet Rd., Hanoi 100000, Vietnam. }}

\section{Introduction}
Sulfation (i.e. metal sulfate formation) of noble metal based catalysts has been a serious problem to automotive emissions aftertreatment systems.\cite{RefWorks:604,RefWorks:217,RefWorks:281,RefWorks:323,RefWorks:298,RefWorks:243} It is well established that Pd is extremely susceptible towards sulfation (i.e., the PdSO$_4$ formation) in the highly oxidizing and sulfating environment typically experienced by the aftertreatment catalysts. Unlike the easily formed sulfate PdSO$_4$ under catalytically relevant conditions, no experimental evidence is available for the existence of PtSO$_4$ under any circumstances.\cite{RefWorks:623} Despite being a member of the same group of the Periodic Table, an intriguing fact of non-existence of PtSO$_4$ remains as a puzzle and an unexplored territory. A question arises why PtSO$_4$ does not exist and what makes PtSO$_4$ different from PdSO$_4$? Answers to these questions may reveal the underlying reason behind the sulfation resistant phenomena of Pt and, in turn, provide some guidance for future design of sulfur resistant catalysts materials.

An experimental investigation based reaction pathway analysis suggested that the PdSO$_4$ formation is primarily due to the interaction between SO$_3$ and metal oxide (i.e., PdO) in the catalytically relevant temperature and pressure conditions. \cite{RefWorks:292} Nevertheless, no PtSO$_4$ formation has been observed under similar experimental conditions.\cite{RefWorks:214,RefWorks:272} A recent first-principles computation based study suggested that the structure of PtSO$_4$ should be similar to that of PdSO$_4$ while assuming a similar nature of metal oxides (i.e. PdO and PtO) of Pd and Pt.\cite{RefWorks:623}  Using first-principles thermodynamics we have recently predicted that the PdSO$_4$ formation is indeed favored even at lower temperature pressure conditions; however, the PtSO$_4$ formation may be favorable only at elevated pressure conditions.\cite{RefWorks:674}  This outcome points out a direction for further investigations of the Pt and Pd sulfates under a wide range of temperature and pressure regimes, for which comprehensive information on the possible structural phases is required. Furthermore, PdSO$_4$ is stable towards decomposition to metal oxide (PdO) and sulfur oxides (SO$_2$/SO$_3$) below $\sim 650^{\circ}$C \cite{RefWorks:228} which suggests that once the stable sulfate is formed, it is difficult to desulfate the catalysts. Unfortunately, such information is missing for PtSO$_4$ and needs an attention.

Our work is premised on the aforementioned mystery of  contrasting behavior of Pt and Pd metals towards sulfation. We extensively explore the possible low-energy structures of the yet-to-be synthesized PtSO$_4$ and the  known PdSO$_4$ using evolutionary algorithm-based method Universal Structure Predictor: Evolutionary Xtalloraphy (USPEX). \cite{RefWorks:627,RefWorks:628} The thermodynamic stability of the predicted low-energy structures are assessed by the evaluation of Gibbs free energy  over a wide temperature-pressure range, fully considering the vibrational contributions calculated within the harmonic approximation. Furthermore, we investigate the stability of the predicted structures towards decomposition to their possible products.  In this work, most notably we predict a tetragonal $P4_2/m$ structure (no. 84) to be the lowest in energy for both PtSO$_4$ and PdSO$_4$. Interestingly, we find that the experimentally-known monoclinic $C2/c$ phase (no. 15) of PdSO$_4$ is energetically competing with the newly identified $P4_2/m$ phase. From free energies calculations, we propose the temperature-pressure phase diagrams for PdSO$_4$ and PtSO$_4$, predicting the stable phases of the sulfates at high temperatures and/or high pressures.
\begin{figure*}[t]
\begin{center}
\includegraphics[width=16.5cm]{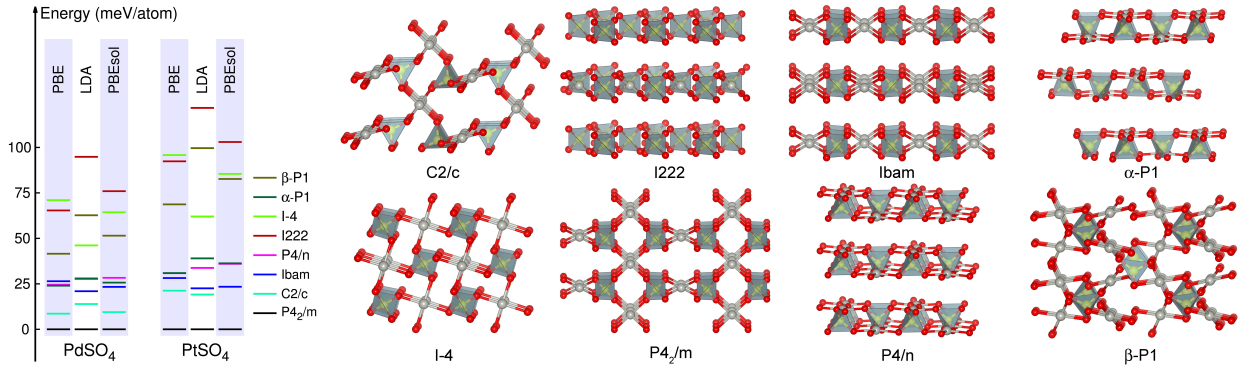}
\caption{Relative energetics and structures of the selected low energy phases of PdSO$_4$ and PtSO$_4$ predicted using USPEX method. Pt (or Pd), S, and O atoms are represented by silver, yellow, and red colors, respectively.}\label{fig:structures}
\label{fig:model}
\end{center}
\end{figure*}

\section{Methods}
Structural phases of compressed matters can now be effectively predicted and discussed at elevated pressures by many  {\it state-of-the-art} computational methods, mostly at the level of first principles. \cite{C4CP04445B} In this work, possible stable structures of PdSO$_4$ and PtSO$_4$ were searched using the evolutionary search technique embodied in USPEX code. \cite{RefWorks:627,RefWorks:628} This code/method, designed to predict the crystal packing from only a knowledge of chemical species, compositions, or the molecular geometries, has met tremendous success in correctly identifying and predicting the crystal structures of various classes of systems (bulk crystals, \cite{RefWorks:628, Refworks:668} nanoclusters,\cite{RefWorks:669} 2D crystals, \cite{RefWorks:670} surfaces,\cite{RefWorks:013} and recently for polymers\cite{RefWorks:671, RefWorks:672}). In this work, we explored of the low-energy configurational spaces of up to four formula units of PdSO$_4$ and PtSO$_4$ per primitive cell, ie., $Z\leq 4$. Structures with $Z > 4$ are not considered, and hence, sets a limitation of this work.

Our first-principles calculations were performed within the framework of density functional theory (DFT) using the projector augmented wave method\cite{RefWorks:336,RefWorks:666} as implemented in Vienna \textit{Ab initio} Simulation Package ({\sc vasp}).\cite{RefWorks:373,RefWorks:375} While the generalized gradient approximation Perdew-Burke-Ernzerhof (PBE)  exchange-correlation (XC) functional was used throughout this work, the energy ordering of the identified structures were confirmed to be invariant with the PBEsol\cite{PBEsol} and the local density approximation (LDA) XC functionals. A basis set of plane waves with kinetic energy up to 600 eV was used to represent the Kohn-Sham orbitals while the Brillouin zones were sampled by well-converged Monkhorst-Pack {\bf k}-point meshes, i.e., no less than $7 \times 7 \times 7$. Convergence in optimizing the structures was assumed when the Hellman-Feynman forces  become less than 0.01 eV/\AA.

We calculated the densities of states (DOS) of the identified structures by the linear tetrahedron method with Bl\"{o}chl corrections. For examining their dynamical stability, the phonon frequency spectra calculated using the finite-displacement approach as implemented in the {\sc phonopy} code. \cite{RefWorks:666a,RefWorks:666b} To establish the stability of the predicted phases at finite temperatures and pressures, relevant thermodynamic properties were evaluated within the harmonic approximation from the computed phonon band spectra. {\sc FullProf} suite\cite{fullprof} was used to simulate the X-ray diffraction patterns.
\section{Results and discussions}

\subsection{Low-energy structures of PdSO$_4$ and PtSO$_4$}
Our evolutionary algorithm based search for low-energy structures of PdSO$_4$ and PtSO$_4$, performed at zero pressure ($P=0$ GPa), returned numerous possible candidates. Eight of them (six $Z=2$ and two $Z=4$ structures), which are lowest in energy for both PtSO$_4$ and PdSO$_4$, and their energetic information are shown in Fig. \ref{fig:structures}. Of the two common thermodynamically most stable structures of these sulfates, one is described by the tetragonal $P4_2/m$ space group (no. 84) while the other belongs to the monoclinic $C2/c$ space group (no. 15). It is worth noting that the $P4_2/m$ structure can also be obtained by substituting Pd/Pt into the Ag sites of the $P\overline{1}$ ($Z=2$) structure of AgSO$_4$. \cite{ANIE:ANIE200906863} On the other hand, the $C2/c$ structure, which was experimentally known for PdSO$_4$,\cite{RefWorks:659} is similar to that discussed earlier by Derzsi {\it et al.}\cite{RefWorks:623} We then found that the $C2/c$ structure of PdSO$_4$ is higher in energy than the $P4_2/m$ structure by $\simeq 8$ meV/atom, falling within the uncertainty of DFT in calculating energies, while for PtSO$_4$, this energy difference is considerably larger, being roughly $20$ meV/atom. The orthorhombic $Ibam$ (no. 72) and the tetragonal $P_4/n$ (no. 85) structures of both PdSO$_4$ and PtSO$_4$ are relatively similar in energy, residing at $\gtrsim 50$ meV/atom above the $P4_2/m$. Two $Z=4$ triclinic structures examined, namely $\alpha$-$P1$ and $\beta$-$P1$ (both no. 1), are about $30-50$ meV/atom above the $P4_2/m$ structure. The last two structures, i.e., $I222$ (orthorhombic, no. 23) and $I\overline{4}$ (tetragonal, no. 82), are about $75-100$ meV/atom higher than the $P4_2/m$ structure. We note that these structures are slightly below the $C2/c$ ($Z=16$) structure recently determined\cite{C2CE26282G} for AgSO$_4$. This energy ordering remains essentially unchanged when PBEsol and LDA were used (also see Fig. \ref{fig:structures}). Crystallographic information of the predicted structures is given in the Supporting Information (see Table S1).

\begin{figure}[t]
\centering
\includegraphics[width=7.5cm]{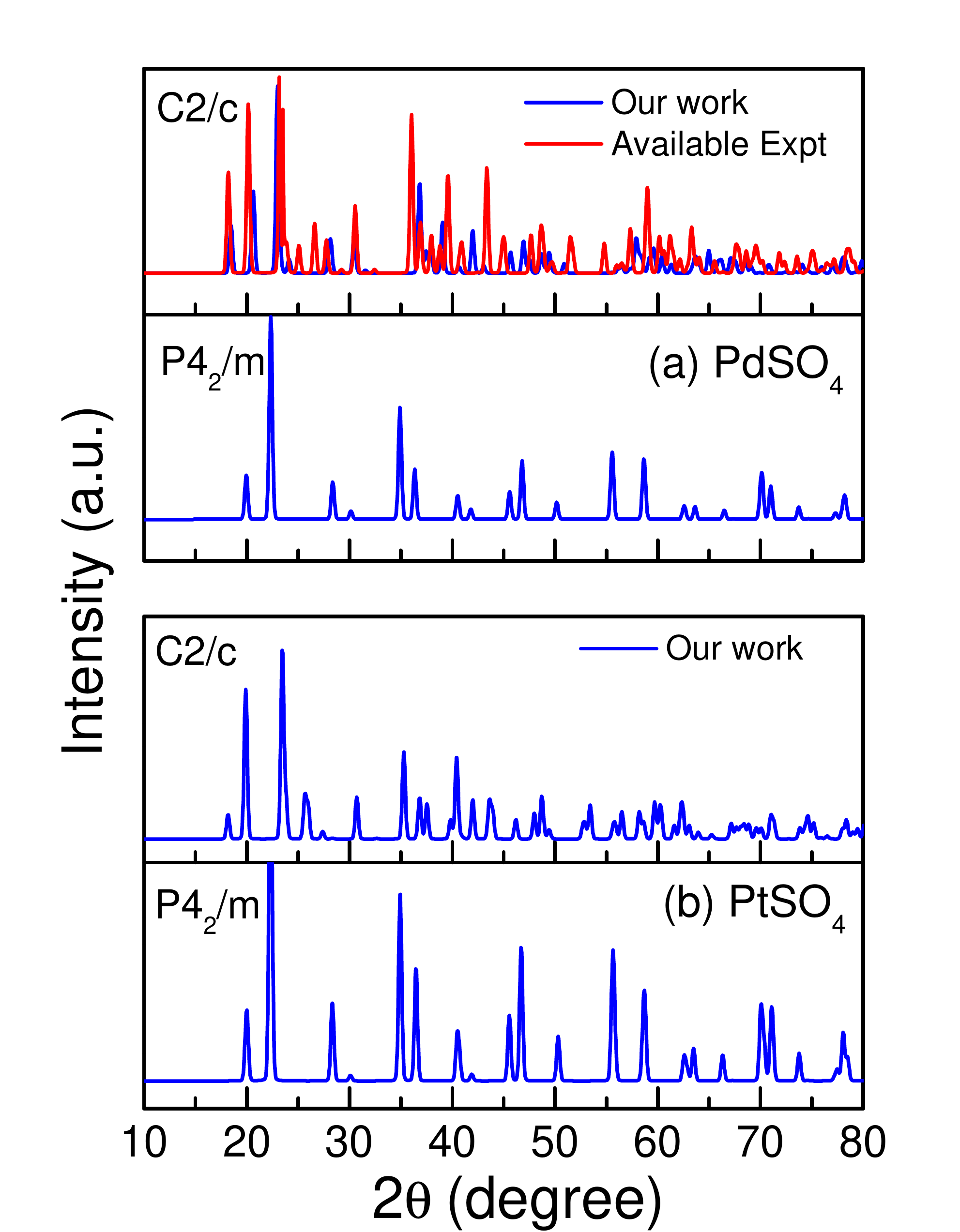}
\caption{\label{fig:XRD} Simulated XRD patterns of two low-energy structures predicted for (a) PdSO$_4$ and (b) PtSO$_4$ at ambient pressure. For PdSO$_4$, the XRD pattern of the monoclinic $C2/c$ is shown using the available experimental data/parameters.\cite{RefWorks:241}  The XRD patterns were simulated at Cu K$\alpha$ with $\lambda=1.54056$\AA.}\label{fig:xrd}
\end{figure}

\begin{figure}[t]
\centering
\includegraphics[width=7.5cm]{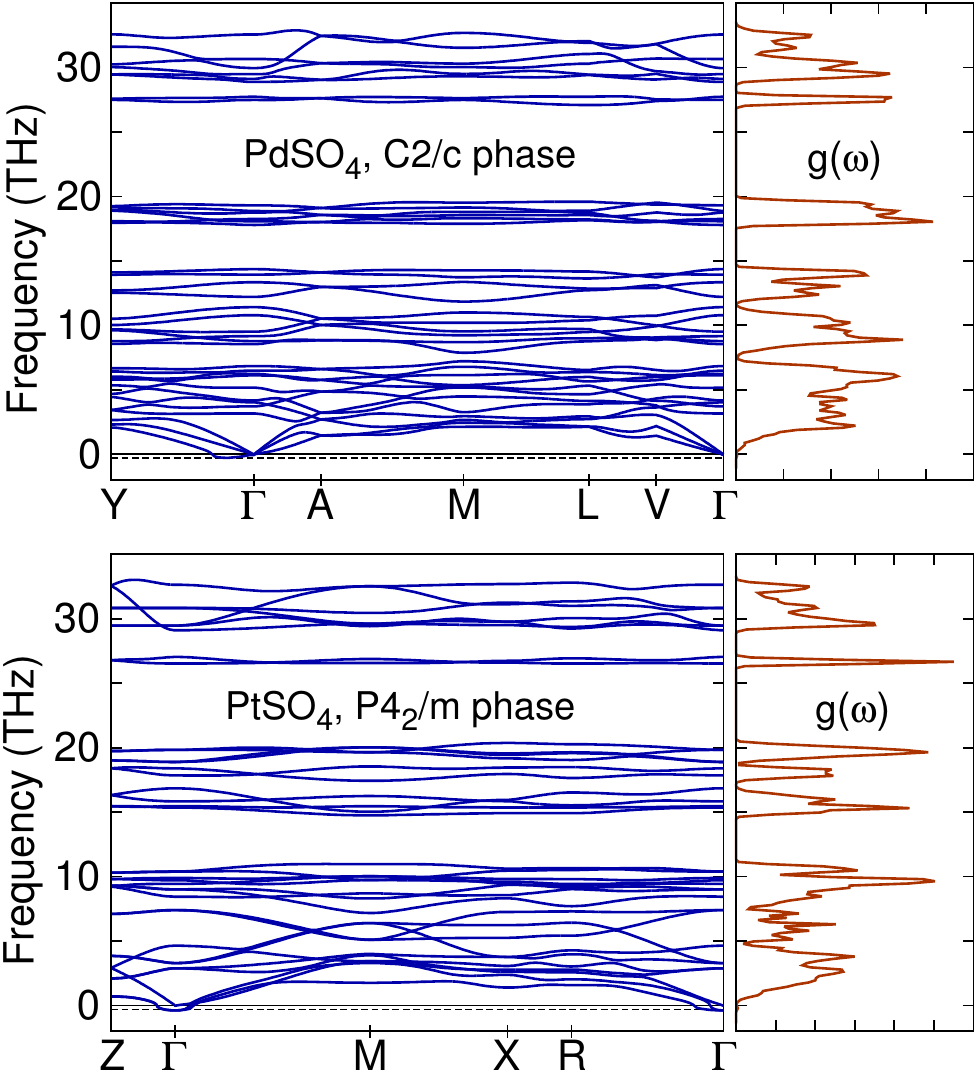}
\caption{\label{fig:phonon} (Color online) Phonon band structures (left) and density of phonon states $g(\omega)$ (right, given in arbitrary units), of the $P4_2/m$ phase of PtSO$_4$ (top panel) and the $C2/c$ phase of PdSO$_4$ (bottom panel) at $P=0$ GPa. For convenience, bands with imaginary frequencies, if any, are shown as those with {\it negative} frequencies. The dotted lines indicate a numerical error of $\sim 0.3$ THz typically resulted from the translational symmetry breaking while calculating the XC energies in the real space.}
\end{figure}

The predicted low-energy structures of both sulfates consist of tetrahedral SO$_4$ groups, where O atoms are associated to four different SO$_4$ tetrahedra coordinating the Pd/Pt atoms on a plane. The local chemistry at the anionic site, the topology and the connectivity of the crystal networks are also qualitatively similar in both sulfates. In general, these eight structures can be classified into two groups. The first structure type, with a non-layered 3-D network, contains oxygen atoms from the SO$_4$ unit which act as a bridge (for example: $P4_2/m$, $C2/c$, $I\overline{4}$, and $\beta$-$P1$ phases) linking metal atoms. The second structure type involves some two-dimensional motifs with isolated layers of Pd/Pt and SO$_4$ tetrahedra. The remaining four structures, i.e., $Ibam$, $P_4/n$, $I222$, and $\alpha$-$P1$ phases, belong to this class.

We further analyzed the selected low-energy structures by simulating the X-ray diffraction (XRD) patterns. In Fig. \ref{fig:xrd} (top two panels), we show the XRD patterns simulated for the $C2/c$ and $P4_2/m$ structures of PdSO$_4$ along with the available experimental XRD data of the $C2/c$ phase.\cite{RefWorks:241} In the bottom two panels of Fig. \ref{fig:xrd}, we show the XRD patterns simulated for our predicted $C2/c$ and $P4_2/m$ phases. Overall, the simulated XRD patterns are in good agreement with the available experimental data.\cite{RefWorks:241} The additional simulated XRD patterns (of the other predicted phases) are given in the Supporting Information S2.
\begin{figure*}[t]
\centering
\includegraphics[scale=0.52]{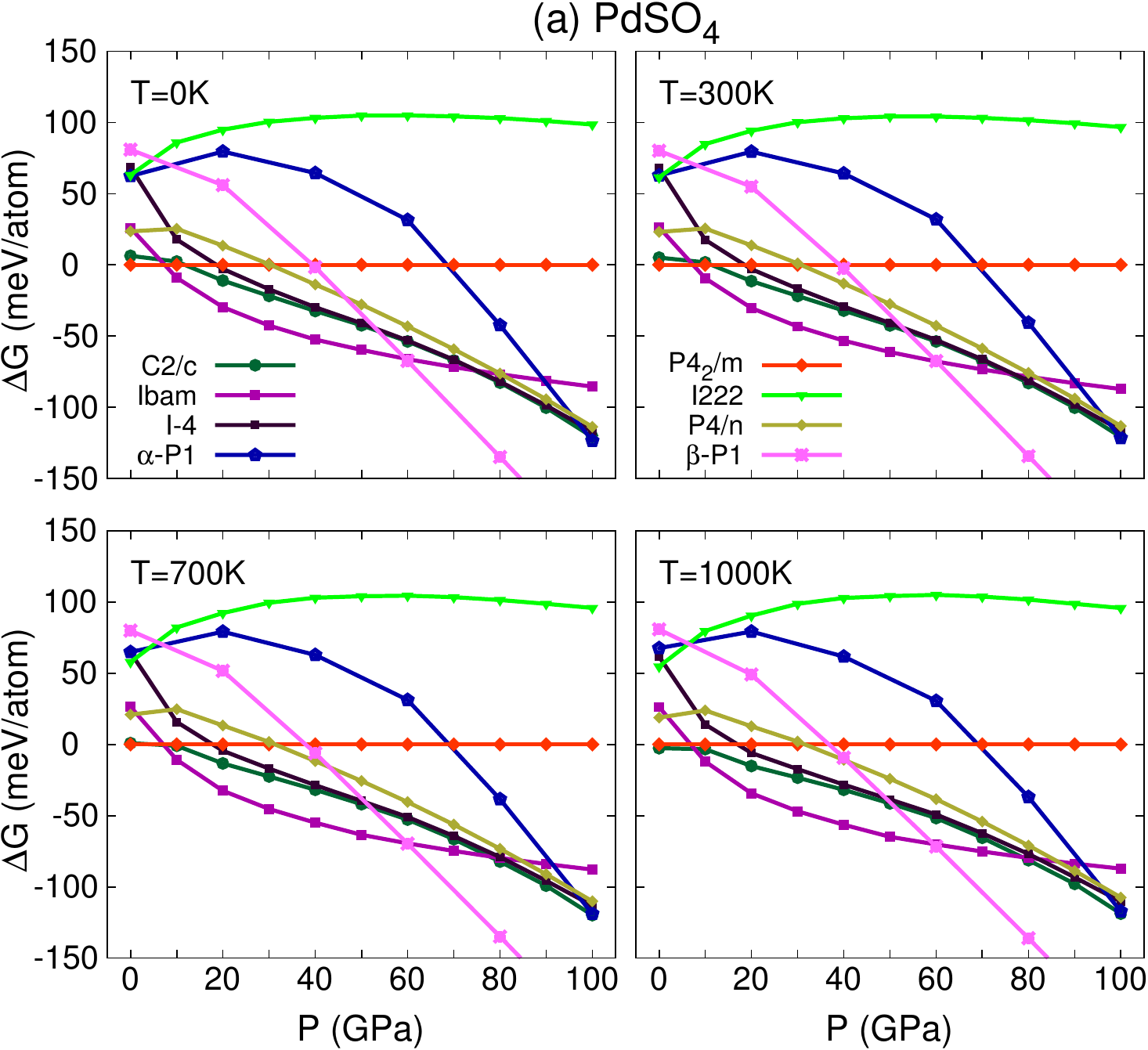}
\includegraphics[scale=0.52]{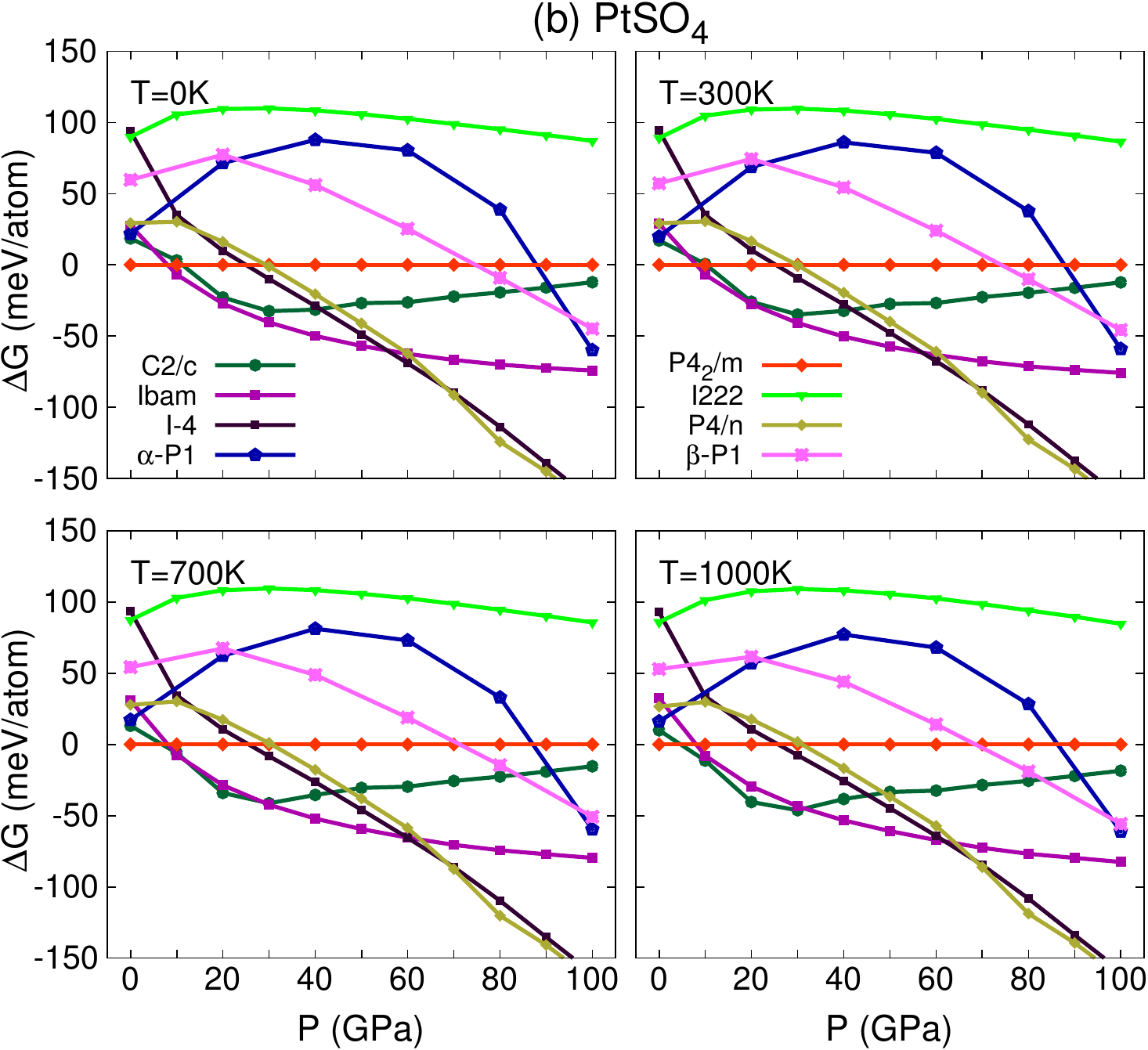}
\caption{\label{fig:freeenergy} Gibbs free energies with entropic contributions, $G$ calculated at $T$ = 0 K, $T$ = 300 K, $T$ = 700 K, and $T$ = 1000 K for the identified low-energy structures of PdSO$_4$ (panel a) and PtSO$_4$ (panel b) are shown as functions of pressure $P$. Data is given by symbols while curves are guides to the eye.}\label{fig:freenergy}
\end{figure*}

\subsection{Dynamical and thermodynamic stabilities}

Next, we examined the dynamical stability of the predicted structures of PtSO$_4$ and PdSO$_4$ using the calculated phonon band structures. No imaginary modes exist throughout the Brillouin zones of these structures, demonstrating that they are dynamically stable.  For illustration, we show in Fig. \ref{fig:phonon} the phonon spectra and the phonon density of states $g(\omega)$ we calculated for the lowest-energy structures of each compound, i.e., the $P4_2/m$ and $C2/c$ structures. Similar information for all other predicted structures can be found in the Supporting Information S3.

The phonon spectra of these structures, calculated at 0 K, allow estimating the vibrational contribution $F_{\rm vib}(T)$ to the Gibbs free energy $G(P,V,T) = E_{\rm DFT} + F_{\rm vib}(T) + PV$ within the harmonic approximation via
\begin{equation}
F_{\rm vib}(T) = rk_{\rm B}T \int_0^\infty d\omega g(\omega) \ln \left[2\sinh\left(\frac{\hbar\omega}{2k_{\rm B}T}\right)\right],
\end{equation}
where, $r$ is number of degrees of freedom in the unit cell, $k_{\rm B}$ is the Boltzmann's constant, $\hbar$ is the reduced Planck's constant, and $g(\omega)$ is the normalized phonon density of state at frequency $\omega$. In addition, the enthalpy $E_{\rm DFT}+PV$ was calculated by slowly optimizing the investigated structures under gradually increasing pressure, starting from $P=0$ GPa. For hard crystalline materials, the method provides a reliable way to access their thermodynamic stability. For hard crystalline materials, this method typically leads to an excellent agreement with experimental data.\cite{RefWorks:673}

The calculated free energies $G(P, V, T)$ are summarized in Fig. \ref{fig:freenergy}, suggesting that the $P4_2/m$ phase of PtSO$_4$ is thermodynamically stable at low pressures. Within this regime, the $C2/c$ structure of PdSO$_4$ (which is experimentally established \cite{RefWorks:659}) is different from the $P4_2/m$ structure by no more than $\pm 2$ meV/atom at low and high temperatures. Therefore, these two phases are considered to coexist at low pressures, The formation of the $C2/c$ phase, which is observed even at low temperatures conditions, may be driven by kinetics, known under the empirical Ostwald's steps rules in crystal nucleation.

Both PdSO$_4$ and PtSO$_4$ undergo several structural phase transitions at elevating pressures. For PdSO$_4$, the orthorhombic $Ibam$ phase is stable between $\sim 10$ and $\sim 60$ GPa before transforming to the triclinic $\beta$-$P1$ phase. The $Ibma$-to-$\beta$-$P1$ phase boundary depend very weakly on temperature. Unlike PdSO$_4$, the $C2/c$ phase of PtSO$_4$ is thermodynamically stable only at high temperature ($\gtrsim 700$ K) and intermediate pressure ($10-30$ GPa) conditions while the $Ibam$ phase is stable at lower temperatures ($\lesssim 700$ K) and elevated pressure ($10-60$ GPa) conditions. The transition between the $Ibam$ phase to the $I\overline{4}$ phase occurs at roughly around $60$ GPa. For both PdSO$_4$ and PtSO$_4$, the $I222$ phase is unstable over the whole range of pressure examined.

\begin{figure}[t]
\centering
\includegraphics[width=8.25cm]{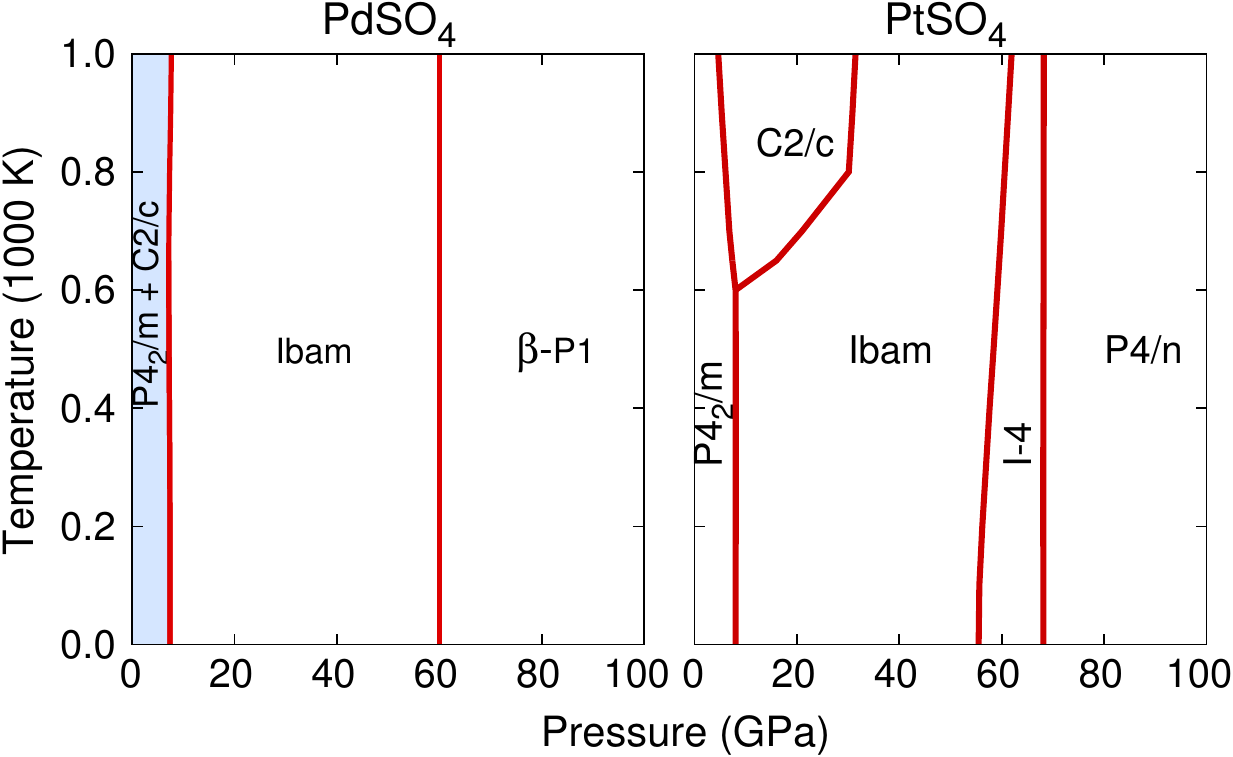}
\caption{\label{fig:phase} Computed phase diagrams of PdSO$_4$ (left panel) and PtSO$_4$ (right panel). Thermodynamically stable phases are shown as indicated by their space group symbols. Shaded area indicates the coexisting regime of both the $P4_2/m$ and $C2/c$ phases of PdSO$_4$.}\label{fig:diagram}
\end{figure}

Using the calculated free energies $G(P,V,T)$ (as shown in Fig. \ref{fig:freenergy}), we constructed the temperature-pressure phase diagrams of both sulfates and show them in Fig. \ref{fig:diagram}. The phase diagrams display a map of the stable phases over the range of $T$-$P$ conditions. Most importantly, we observed that for PdSO$_4$, both the tetragonal $P4_2/m$ and the monoclinic $C2/c$ phases coexist at atmospheric pressures while for PtSO$_4$, the $P4_2/m$ phase is the sole candidate at the same conditions. Furthermore, the $Ibam$ phase dominates the $10-60$ GPa region for both cases, which could be an interest of exploration for the high pressure applications.

While experimental studies suggest that PdSO$_4$ decomposes above $\sim 900$ K\cite{RefWorks:292}, estimation of free energy of reaction ($\Delta G$) allows us to evaluate the thermodynamic stability (reaction feasibility) of the compound towards the decomposition into possible products. The feasibility of a reaction depends on the sign of $\Delta G$, which is equal to $\Delta H - T\Delta S$, where $\Delta H$ is the change in enthalpy and $\Delta S$ is the change in entropy. The $\Delta G$ of the reaction can be expressed as:
\begin{equation}
\Delta G = \sum_{i=1}^{n} G_{\rm products} - \sum_{i=1}^{n} G_{\rm reactants}.
\end{equation}

In this work, we considered the decomposition reaction of Pd(or Pt)SO$_4$ towards their respective most stable metal oxides and sulfur oxide species [i.e. Pd(or Pt)SO$_4$ $\rightarrow$ Pd(or Pt)O + SO$_3$]. For example, the computed $\Delta G$ values at 300 K were $\sim -60$ kJ/mol and $\sim -40$ kJ/mol for PdSO$_4$ and PtSO$_4$, respectively. Furthermore, we evaluated the free energy of the decomposition of the sulfates to their respective elemental species (i.e. Pd (or Pt) SO$_4$ $\rightarrow$ Pd(or Pt) + S + 2O$_2$). The computed $\Delta G$ values were in the range of $\sim - 500$ kJ/mol at 300 K. Results suggest that PdSO$_4$ is stable towards decomposition to PdO and SO$_3$ below 775K whereas PtSO$_4$ stability towards PtO and SO$_3$ remains below 650K. Similarly, PdSO$_4$ is stable towards the decomposition to the elemental components below 870K  whereas PtSO$_4$ is stable below 800K. Furthermore, we computed the dG for the reaction Pd(or Pt)SO$_4$ $\rightarrow$ Pd (or Pt)S + 2O$_2$, which further supports the stability of the sulfates in realistic temperature ($<700$K) and pressure conditions. In general, our calculations show that PdSO$_4$ is more stable than PtSO$_4$ towards decomposition for a particular temperature. Our results are in good agreement, given the computational error range in energetics, with the available experimental results of PdSO$_4$ decomposition stability. Furthermore, synthesis of PtSO$_4$ seems feasible in the future given the kinetic barriers are easy enough to cross. The free energy ($\Delta G$) versus temperature ($T$) plot is provided in Supporting Information (S4, Figure S3).

To further confirm whether these sulfates are stable or not with respect to the pool of all possible product species, a linear programming (LP) algorithm\cite{RefWorks:676,RefWorks:677} has been employed. Here, a Pd(or Pt)SO$_4$ compound is considered to be stable when $\Delta E$ (the DFT energy relative to the best outcome from the LP) is negative. The energy difference, $\Delta E$, can thus be written as
\begin{equation}
\Delta E = {\rm Pd(or~Pt)SO}_{4} - \min \sum_{i=1}^{n} c_{i} P_{i},
\end{equation}
where $P_{i}$ represents all the possible stable chemical species (i.e. for PdSO$_4$: Pd, PdO, PdS, SO$_3$, SO$_2$, SO, S, and O$_2$; for PtSO$_4$: Pt, PtO$_2$, PtO, PtS, SO$_3$, SO$_2$, SO, S, and O$_2$). For example, the equation for PdSO$_4$ becomes
$\Delta E = {\rm PdSO}_4 - \min(c_{1} {\rm Pd}_{a_{1}} + c_{2} {\rm Pd}_{a_{2}}{\rm O}_{o_{2}} + c_{3} {\rm Pd}_{b_{3}}{\rm S}_{3o_{3}} + c_{4} {\rm S}_{b_{4}}{\rm O}_{3o_4}$ + $c_5 {\rm S}_{b_5}{\rm O}_{2o_5} + c_6 {\rm S}_{b_6} {\rm O}_{o_6}+ c_7 {\rm S}_{b_7} + {\rm c}_8 {\rm O}_{2o_8})$.
Then, the LP problem is solved with the constraints
\begin{equation}
\sum_{i} a_{i}c_{i} =1,~\sum_{i} b_{i}c_{i} =1,~ {\rm and}~ \sum_{i} o_{i}c_{i} =4,
\end{equation}
where $a_{i}$, $b_{i}$, and $o_{i}$ represent Pt(or Pd), S, and O content of a species, respectively. Above constrains ensure the correct stoichiometry of  Pd(or Pt)SO$_4$ and with
\begin{equation}
 c_{i} \geq1,
 \end{equation}
 which warrants that only the references containing Pt(or Pd), S, or O are taken into account. With all DFT computed energies of the species, we obtained all the optimized $c_{i}$ and $\Delta E$ for each case. Consistent with our free energy of reaction analysis, negative $\Delta E$ values (i.e. $-0.87$ eV and $-0.70$ eV for PdSO$_4$ and PtSO$_4$, respectively) were obtained, which confirmed the stability of the sulfates. Interestingly, we obtained mono-metallic oxides (PdO and PtO in the case of PdSO$_4$ and PtSO$_4$, respectively) and SO$_3$ as possible decomposition products,consistent with our reaction free energy analysis, and unity (as expected) for all $c_{i}$ values.

\subsection{Electronic structures}
We  investigated the electronic structures of all low-energy phases of PdSO$_4$ and PtSO$_4$ by computing the total density of states. Overall, our results show no significantly different behavior between the phases of both sulfates. For better understanding the DOS  can be divided into three main groups. First, the lower valence bands (between $-2$ eV and $-4$ eV) originate due to mixing of the valence $d$ and $p$ states of Pd(or Pt) and the O atoms. In this region, Pd(or Pt) ($d$) bands are found to be highly resonant with the O ($p$) bands. We also noticed that some pronounced mixing between the segments of O ($p$) bands lying above and below the valence Pd(or Pt) ($d$) bands. Second, in the vicinity of the Fermi level the valence-band maximum are dominated by Pd(or Pt) ($d$) states. Third, the bottom of the conduction band consists $3p$ states of S and O ($2p$) states. Further detail can be found in the Supporting  Information S5.

\section{Conclusions}
In summary, we explored the mystery related to nonexistence of PtSO$_4$ using first-principles thermodynamics combined with the evolutionary algorithms based method. Our approach is validated by also studying the experimentally known phases of PdSO$_4$. Many low-energy structures are predicted and analyzed for the stability in a wide range of temperature and pressure conditions. At low pressures, we identify a tetragonal $P4_2/m$ structure (of the AgSO$_4$ type) which appears to be the thermodynamically most stable phase of PtSO$_4$. In case of PdSO$_4$, this phase is predicted to coexist with the experimentally known $C2/c$ phase.  These sulfates are also predicted to undergo several phase transitions at elevated temperatures and/or pressures. Based on the computed Gibbs free energies, we constructed phase diagrams which provide such the reliable information about the phases stability, the phase transition, and their boundaries up to $100$ GPa and $1000$ K. The phase diagrams confirmed the existence of experimentally observed monoclinic $C2/c$ phase of PdSO$_4$ at the ambient conditions; however, this phase may not be seen in the case of PtSO$_4$ in similar conditions. Nonetheless, $Ibam$ phase remains one of the promising stable phase for both cases at high pressure conditions. Both sulfates were stable towards decomposition to their possible products well above the room temperature, which also suggests the possibility of PtSO$_4$ synthesis in the future. In general, we provide a detailed information on the phases and their stability of PdSO$_4$ and PtSO$_4$ which can be helpful to understand the sulfating nature of Pd and design/scan promising new sulfur resistant materials.

\textbf{Acknowledgement}
The authors thank Prof. Rampi Ramprasad for valuable discussions. HNS acknowledges the United States Environmental Protection Agency (EPA) STAR graduate fellowship, fellowship number FP917501, for funding support. Its contents are solely the responsibility of the fellow and do not necessarily represent the official views of the EPA. Finally, Authors acknowledge Materials Design\textsuperscript{\textregistered} for MedeA\textsuperscript{\textregistered} software and thank School of Engineering, Univ. of Connecticut for Hornet supercomputer access.

\textbf{Supplementary Information}

Electronic Supplementary Information (ESI) available: Structural information of PtSO$_4$ and PdSO$_4$ are shown in S1, XRD patterns of the predicted structures are shown in S2, the phonon density of states are shown in S3, free energy diagram is given in S4, and electronic density of states are shown in S5 of the Supplementary Information. 

$^*$Corresponding author: \url{huan.tran@uconn.edu}




\providecommand*{\mcitethebibliography}{\thebibliography}
\csname @ifundefined\endcsname{endmcitethebibliography}
{\let\endmcitethebibliography\endthebibliography}{}

\end{document}